\documentclass{article}

\usepackage{arxiv}
\usepackage[utf8]{inputenc}

\usepackage{longtable}
\usepackage{authblk}
\usepackage[caption = false]{subfig}
\usepackage{graphicx}
\usepackage{amssymb}
\usepackage{amsmath}
\usepackage{amsthm}
\usepackage[noend]{algpseudocode}
\usepackage{algorithm}
\usepackage{titletoc}
\usepackage{titlesec}
\usepackage[utf8]{inputenc}
\usepackage{enumerate}
\usepackage{hhline}

\makeatletter
\renewcommand{\Function}[2]{%
	\csname ALG@cmd@\ALG@L @Function\endcsname{#1}{#2}%
	\def\jayden@currentfunction{#1}%
}
\newcommand{\funclabel}[1]{%
	\@bsphack
	\protected@write\@auxout{}{%
		\string\newlabel{#1}{{\jayden@currentfunction}{\thepage}}%
	}%
	\@esphack
}
\makeatother

\newcommand{\stair}{{\rm STAIR}}

\usepackage[english]{babel}

\newcommand{\dfnn}[2]{ \textbf{\emph{[#1]}} {#2}}

\makeatletter
\def\BState{\State\hskip-\ALG@thistlm}
\makeatother

\title{StairDag: Cross-DAG Validation For Scalable BFT Consensus}

\author{Quan Nguyen, Andre Cronje, Michael Kong, Alex Kampa, George Samman}
\affil{FANTOM}

\begin{document}
\maketitle

\begin{abstract}
This paper introduces a new consensus protocol, so-called \emph{\stair}, for fast consensus in DAG-based trustless system. 
In \stair, we propose a new approach to creating local block DAG, namely \emph{x-DAG} (cross-DAG), on each node. \emph{\stair} protocol is based on our Proof-of-Stake StakeDag framework~\cite{stakedag} that distinguishes participants into users and validators by their stake. Both users and validators can create and validate event blocks. 
Unlike StakeDag's DAG, x-DAG ensures that each new block has to have parent blocks from both Users and Validators to achieve more safety and liveness. Our protocol leverages a pool of validators to expose more validating power to new blocks for faster consensus in a leaderless asynchronous system.
Further, our framework allows participants to join as observers / monitors, who can retrieve DAG for post-validation, but do not participate in onchain validation.

\end{abstract}


\newpage
\pagenumbering{arabic} 
\tableofcontents 
\newpage

\section{Introduction}\label{ch:intro}

There have been rising interests in cryptocurrencies and distributed ledger technologies since the introduction of Bitcoin in late 2008. The underlying blockchain technology (BCT) backing the decentralization principle have attracted numerous applications across different domains from financial, healthcare to logistics. 
Blockchains provide trustworthy immutability and transparency of blocks and are emerging as a promising solution for building trustless systems, which ensure trust in the system with no assumed trust of its participants.
Public blockchain systems support third-party auditing and some of them support a high level of anonymity. 

Blockchain is generally a distributed database system. Each blockchain is a database storing all transactions within the network and is replicated to all participating nodes. A distributed consensus protocol running on each node to guarantee the consistency of the replicas so as to maintain a common transaction ledger.
In a distributed database system, reliability of the system is addressed as \emph{Byzantine} fault tolerance (BFT)~\cite{Lamport82} which is tolerant up to one-third of the participants in failure. Consensus algorithms, which ensure the integrity of transactions over the distributed network, are equivalent to the proof of BFT~\cite{randomized03, paxos01}. 
In practical BFT (pBFT) can reach a consensus for a block once the block is shared with other participants and the share information is further shared with others \cite{zyzzyva07, honey16}.

\subsection{Overview of blockchains}
Bitcoin and blockchain technologies have enabled numerous opportunities for business and innovation, facilitating highly trustworthy, append-only, transparent public distributed ledgers.
Despite of the great success, blockchain systems are still facing some limitations. 
Recent advances in consensus algorithms \cite{ppcoin12, dpos14, dagcoin15, algorand17, sompolinsky2016spectre, PHANTOM08} have improved the consensus confirmation time and power consumption over blockchain-powered distributed ledgers. 

\emph{Proof of Work} (PoW): is the most used model since introduced in the Bitcoin's Nakamoto consensus protocol~\cite{bitcoin08}. 
Under PoW, validators are randomly selected based on their computation power. PoW protocol requires exhausive computational work and so electricity from participants for block generation, and a long time for transaction confirmation.

\emph{Proof Of Stake} (PoS): leverages participants' stakes for selecting the creator of the next block~\cite{ppcoin12,dpos14}. Validators have voting power proportional to their stake. PoS requires less power consumption than PoW. 
PoS is also more secure than PoW, as the stake of a participant is voided and burnt if found dishonest.

\emph{DAG} (Directed Acyclic Graph): DAG based currency was first introduced in DagCoin paper~\cite{dagcoin15}. DAG technology becomes a promising alternative that allows cryptocurrencies to function similarly to those that utilize blockchain technology without the need for blocks and miners. 
DAG-based approaches have recently emerged as a promising alternative than the proof of work (PoW) for consensus. 
These approaches utilize directed acyclic graphs (DAG)~\cite{dagcoin15, sompolinsky2016spectre, PHANTOM08, PARSEC18, conflux18} to facilitate consensus. 
Examples of DAG-based consensus algorithms include Tangle~\cite{tangle17}, Byteball~\cite{byteball16}, and Hashgraph~\cite{hashgraph16}.

\emph{PoS + DAG}: StakeDag protocol~\cite{stakedag} presents a general model that integrates Proof of Stake model into DAG-based consensus protocols.
Like its predecessors~\cite{lachesis01,fantom18}, the StakeDag consensus protocol generate each block asynchronously to build the DAGs (S-OPERA chain) from User and Validator blocks. Consensus on a block is computed from the gained validating power of Users and Validators on the block.
Like ONLAY framework~\cite{onlay19}, StakeDag's $S_\phi$ protocol uses the concepts of graph layering and hierarchical graphs on the DAGs. Then assigned layers are used to achieve deterministic topological ordering of finalized event blocks in an asynchronous leaderless DAG-based system.

\subsection{\stair\ Framework}

In this paper, we propose a new consensus protocol, namely \stair\ (\underline{S}take-based \underline{T}rustless \underline{A}synchronous \underline{I}mmutable \underline{R}eplicator), for PoS DAG-based distributed systems.
\stair\ protocol aims for pBFT consensus in a trustless system using asynchronous event transmission to replicate leaderless, scalable, asynchronous DAG to all participant nodes. 

We distinguish validating nodes into users and validators based on their stake. Validators, who have a high amount of stake, have more validating power and thus can gain more validation rewards.
In \stair, we introduce a new model of DAG, called x-DAG, for better cross validation amongst the participants. Each new event block in x-DAG has parent blocks from User and Validator blocks. Intuitively, User block can get more validating power from Validator's parent block; and Validator must validate a User block (parent block) in order to create a new Validator block.
Thus, x-Dag boosts up the cross validation between Users and Validators to reach faster consensus. 

Further, we propose a new participant type, namely Observer, who ensures the consistency of the global ledger as well as the expected honesty of validating nodes. Observer nodes do not participate in online voting for consensus, but rather they do post-validation and detect any malicious nodes and invalid transactions.

We then present a simple staking model for the new PoS-DAG protocol. The new staking model gives an intuitive way to compute and update validating power from participant's stake.

Like StakeDag, \stair\ framework compute asynchronous partially ordered sets with logical time ordering instead of blockchains. \stair\ offers a new practical alternative framework for distributed ledgers.

The main concepts of \stair\ are given as follows:
\begin{description}
	\item [$-$ Stake] is the amount of tokens each node posesses in their staking deposit. This value decides the validating power a node can have.
	\item [$-$ User node] A user node has a small amount of stake (e.g., containing from 1 to $U$ -1 = 999 FTM tokens).
	\item [$-$ Validator node] A validator node has large amount of stake (from $U$ = 1,000 FTM tokens).
	\item [$-$ Observer node] An observer node does not require to have any stake (zero stake).
	\item[$-$ Event block] An immutable set of transactions created by a node. Event block includes signature, timestamp, transaction records and referencing hashes to previous (parent) blocks. Blocks are classified by Black blocks (created by a low trust node --- User) and Red blocks (created by a high trust node --- Validator).	
	\item [$-$ x-DAG] is the local view of the weighted Directed Acyclic Graph (DAG) held by each node. This local view is used to determine consensus.	
	\item [$-$ Saga point] is a reward point to a validating node for successfully validating a block that is eventually finalized.
\end{description}

\subsection{Paper structure}

The rest of this paper is organised as follows. 
Section~\ref{se:related} gives the related work. 
Section~\ref{se:model} presents our general model of Proof of Stake DAG-based consensus protocols.
Section~\ref{se:staking} 
introduces a staking model that is used for our \stair\ consensus protocol.
Section~\ref{se:discuss} gives some discussions about our Proof of Stake \stair\ protocol, such as fairness and security.
Section~\ref{se:con} concludes.

\section{Related work}\label{se:related}

Blockchain technology is the future platform for distributed ledger technology (DLT). Blockchain stores transactions into blocks that are linked together. The blocks and their chain are immutable and therefore serving as a proof of existence of a transaction. BCT and its decentralization principle has seen a large applications across many sectors such as finance, energy, public services and sharing platforms. Public, permissionless blockchain relies on no central authority. Instead, consensus among users is paramount to guarantee the security and the sustainability of the system.

\textbf{Proof of Work} 
Proof of Work (PoW) is the consensus protocol of the Bitcoin~\cite{bitcoin08}.
PoW relies on user's computational power to solve a cryptographic puzzle to achieve consensus and integrity of data stored in the chain. Nodes validate transactions in blocks with an incentive to gain block reward and transaction fees associated to the transactions. PoW comes together with an enormous energy demand. 
Several alternatives to Proof-of-Work have been proposed.

\subsection{Proof of Stake}

\emph{Proof of Stake}(PoS) relies on a lottery like system to select the next leader or block submitter. Instead of using puzzle solving, PoS elects the next node based on the amount of stake or the coin age that node has. The elected participant can issue the next block and then is rewarded with transaction fees from participants whose data are included. 
The underlying assumption is that: stakeholders are incentivized to act in its interests, so as to preserve the system.

There are two major types of PoS. First, chain-based PoS~\cite{pass2017fruitchains}, such as Peercoin~\cite{king2012ppcoin}, Blackcoin~\cite{vasin2014blackcoin}, and Iddo Bentov’s work~\cite{bentov2016}, uses chain of blocks like in PoW but selects stakeholders randomly based on their stake to create new blocks. Second, BFT-based PoS~\cite{kwon2014tendermint,algorand16} is based on BFT consensus algorithms such as pBFT~\cite{Castro99}.

PoS is more advantangous than PoW because it only requires cheap hardware and every validator can submit blocks. 
PoS approach reduces the energy demand and is more secure than PoW.
But PoS has new issues arise that were not present in PoW-based blockchains, such as Grinding attack and Nothing at stake attack.

{\bf Delegated Proof of Stake}
\emph{Delegated Proof of Stake} (DPoS) consensus protocols are introduced, such as Lisk, EOS~\cite{EOS}, Steem~\cite{Steem}, BitShares~\cite{bitshares} and Ark~\cite{Ark}. 
In DPoS, users can \emph{vote} to select \emph{witnesses}, who have the right to create blocks, validate transactions, and earn rewards. Further, users can also \emph{delegate} their voting power to other users to vote for witnesses on their behalf. Users can also vote to remove a top tier witness who has lost their trust. Thus, the potential loss of income and reputation is the main incentive against malicious behavior in DPoS.
Users in DPoS systems also vote for a group of \emph{delegates}, who are responsible for the governance and performance of the network.
For example, they can propose to change block size, or the reward a witness can earn from validating a block. The proposed changes will be voted by the system's users. But the delegates cannot do transaction validation and block generation.

DPoS brings various benefits:
(1) faster than traditional PoW and PoS systems;
(2) enhance security and integrity of the blockchains as each user has an incentive to perform their role honestly.
(3) normal hardware is sufficient to join the network and become a user, witness, or delegate.
(4) more energy efficient than PoW.

{\bf Leasing Proof Of Stake} Another type of PoS is Leasing Proof Of Stake (LPoS). Like DPoS, LPoS allows users to vote for a delegate that will maintain the integrity of the system. Users in a LPoS can lease out their coins and share the rewards gained by validating a block.

{\bf Proof of Authority} Another successor of PoS is \emph{Proof of Authority} (PoA). In PoA, the reputation of the validator acts as the stake~\cite{ProofofAuth}. PoA runs by a set of validators in permissioned system. It gains higher throughputby reducing the number of messages sent between the validators. Reputation is difficult to regain once lost and thus is a better choice for ``stake''.

Our new \stair\ consensus protocol is a public permissionless Proof of Stake protocol. Like StakeDag, it allows participants to delegate their stake to a node to increase validating power of a node and share the validation rewards. 
Interested readers may see more details in our previous paper~\cite{stakedag}. 

\subsection{DAG-based approaches}

The notion of a \emph{DAG} (directed acyclic graph) was first coined by DagCoin~\cite{dagcoin15}. DAG-based approaches are a promising alternative to the PoW and PoS blockchains.  Unlike a blockchain, DAG-based system facilitate consensus while achieving horizontal scalability. DAG technology has been adopted in numerous systems.
Examples of DAG-based approaches include Tangle~\cite{tangle17}, Byteball~\cite{byteball16}, 
Hashgraph~\cite{hashgraph16}, 
RaiBlocks~\cite{raiblock17},
Phantom~\cite{PHANTOM08} ,
Spectre~\cite{sompolinsky2016spectre},
Conflux~\cite{conflux18},
Parsec~\cite{PARSEC18} and Blockmania~\cite{Blockmania18}
.

In this paper, our \stair\ protocol is different from the DAG-based previous work. It is based on our StakeDag ~\cite{stakedag}, which is PoS+DAG approach. \stair\ uses Proof of Stake in asynchronous permissionless BFT systems.

\section{The Framework}\label{se:model}

In this section, we present our \stair\ framework to achieve Proof of Stake DAG-based consensus in trustless systems. \stair\ aims for a public blockchain in which any participant can join the network.
Like StakeDag, the new consensus protocol consists of nodes that vary in their amount of stake. Participants' stakes are paramount to a trust indicator.

\subsection{Stake}
Each participant node of the system is associated with an account. The \emph{stake} of a participant is truly their account balance, which is the number of tokens that was purchased, accumulated and/or delegated from other account.
Each participant has an amount of stake, denoted by $w_i$.

Each participant with a non-positive balance can join as part of the system. 
In order to participate in generating and validating event blocks, a node needs to have a positive $w_i$. 
In our \stair\ model, a participant with a zero balance can still participate in the network. However, this zero-balance node can neither perform block creation nor block validation.

\subsection{Validating Power and Roles}

In a general model of trustless system, we distinguish active participants in the consensus protocol into users and validators, based on their stake or trust. 
Users have a default low score (less than some amount of stake) of trust.
Validators are high profile with a high trust score.

Every node has a stake value of $t_i$. This stake value is then used to determine their trust or validating power in validation of event blocks.
Every node in \stair\ protocol with a positive stake can create new event blocks and can gossip with all other nodes  and validate the event blocks. All these active nodes, so-called validating nodes, can validate the event blocks. Their validating power $w_i$ is determined by their stake.

\subsection{Users, Validators and Observers}
For an asynchronous DAG-based system, \stair\ supports two types of validating nodes: users and validators.
Each validating node can create new event blocks. Generation of a new even block indicates that the new block and all of its ancestors have been validated by the creator node.
Users have less than $U$ FTM tokens, where $U$ is a predefined threshold value, say $U$ = 1,000. Each validator must have more than $U$ tokens.

\begin{figure}[ht]
	\centering
	\includegraphics[width=0.5\linewidth]{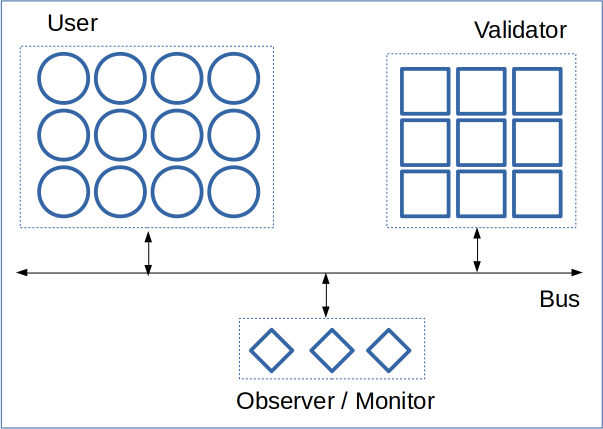}
	\caption{Roles in \stair\ network}
	\label{fig:architecture}
\end{figure}
Besides validating nodes, \stair\ framework allows another type of participants --- observers, who can join the network, but do not act as validating nodes. Observers are not required to have any stake to join the network and thus they cannot generate nor perform online voting (online validation). But observers can do post-validation. This is to encourage the community to join the checks-and-balances of the public network.

There are three general types of nodes that we are considered in our new protocol. They are described as follows.
\begin{itemize}
	\item{\bf{Validators}} {Validator node can create and validate event blocks. They can submit new transactions via their new event blocks. To be a validator, a node needs to have more than $U$ = 1,000 FTM tokens. The validating power of a Validator node is computed by $w_i = U \times \lfloor \frac{t_i}{U} \rfloor$.
	}
	\item{\bf{Users}} {User node can create and validate event blocks. They can submit new transactions via their new event blocks. To be a user, a node needs to have more than 1 FTM token. The number of tokens of a User is from 1 to $U$-1 = 999 FTM tokens. All users have the same validating power of $w_i$ = 1.
	}	
	\item{\bf{Observers:} } {Observer node can retrieve even blocks and do post-validation. Observers cannot perform onchain voting (validation). Observers have zero FTM tokens and thus zero validating power (e.g., $w_i$ = 0).
	}
\end{itemize}

The value of $U$ is system configurable, which is determined and updated by the community. Initially, $U$ can be set to a high value, but can be reduced later on as the token value will increase.
Note that, there is an even simpler model in which all validators have the same validating power of $w_i$ = $U$. But that simplicity comes with a cost as it cannot distinguish validators of different levels of contributions.

\subsection{Event Block Creation}

In order to generate a new event block, a node $n_i$ selects $k$ top event block(s) of other nodes as the block's parents.
Prior to creating a new event block, the node will first validate its current top event block and the selected top event block(s) from other node(s).
 
Like StakeDag, each event block in \stair\ has $k$ references to other event blocks. A validating node (either users and validators) can choose $k$-1 top event block(s) of another validating node(s) as other-parent references to the new event block. Unlike StakeDag, \stair\ requires one of the $k$-1 other-ref blocks is from a creator of an opposite type. 

For a node $n_i$, the $k$ references of a new event block must satisfy the following conditions:
\begin{enumerate}	
	\item One of the $k$ is a self-ref (or self-parent) referencing to an event block of the same node.
	\item The other $k$-1 references, which are also known as other-parent (or other-ref) references, refer to $k$-1 top event blocks on other nodes.
	\item If $n_i$ is a User, then one of the $k$-1 other-ref must be from a Validator block. If $n_i$ is a Validator, then one of the $k$-1 other-ref must be a User block.
\end{enumerate}

\subsection{x-DAG: A Weighted DAG}

We introduce x-DAG, which is a weighted DAG stored as a local history on each node. In our new protocol, each event block has one self-parent reference to the top event block of the same creator, and $k$-1 other-parent references to the top blocks of other nodes.

\begin{figure}[ht]
	\centering
	\includegraphics[width=0.4\linewidth]{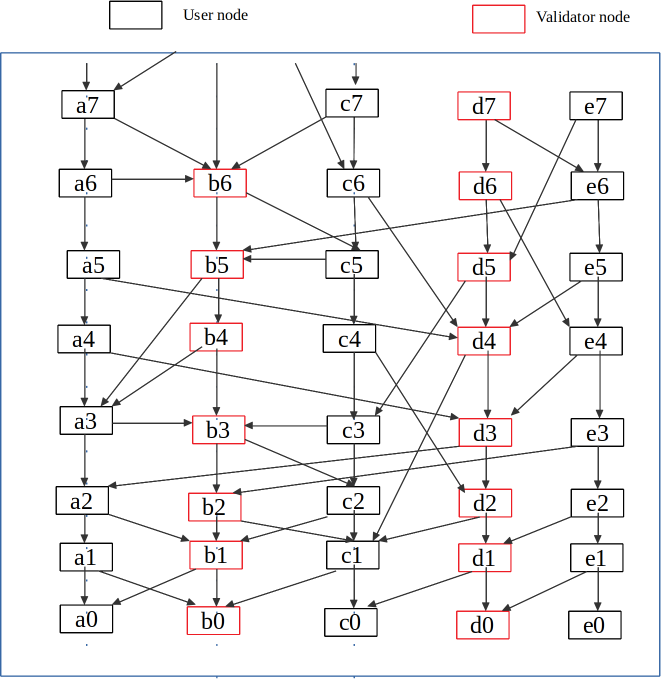}	
	\caption{Examples of x-DAG with users and validators}
	\label{fig:stakedag-ex}
\end{figure}
\stair\ protocol maintains the DAG structure x-DAG, which is based on the concept of S-OPERA chain in our StakeDag protocol~\cite{stakedag}. 
x-Dag chain is a weighted directed acyclic graph $G$=($V$,$E$), where $V$ is the set of event blocks, $E$ is the set of edges. Each vertex (or event block) is associated with a \emph{validation score}, which is the total weights of the roots reachable from it. 
When a block becomes a root, it is assigned a \emph{weight}, which is the same with the validating power $p_i$ of the creator node.

Figure~\ref{fig:stakedag-ex} gives an example of x-DAG, which is a weighted acyclic directed graph stored as the local view of each node. There are five nodes in the example, three of which are normal users with validating power of 1, and the rest are validators. Each event block has $k$=2 references. User blocks are colored Black and Validator blocks are in Red. Each Red block has one Red parent (same creator) and one Black parent (created by a User). Each Black block has one Black parent (same creator) and one Red parent (created by a Validator). As depicted, the example x-DAG is a RedBlack k-ary (directed acyclic) graph.

Let $\mathcal{W}$ be the total validating power $p_i$ of all nodes. 
For consensus, the algorithm examines whether an event block has a validation score of at least $2\mathcal{W}/3$. A validation score of $2\mathcal{W}/3$ means the event block has been validated by more than two-thirds of total validating power in the S-OPERA chain. 

\subsection{Saga points: Validation performance}
To measure the performance of participants, we propose a new notion, called Saga points (Sp). The Saga points or simply 'points' of a node $n_i$, is denoted by $\alpha_i$. As an example, Sp point is defined as the number of own events created by $n_i$ that has a parent (other-ref) as a finalized event (Atropos).

In \stair, we propose to use Saga points to scale the final value of validating power, say $w'_i = \alpha_i . w_i$.
Saga point is perfectly a good evidence for the validation achievement made by a node. Saga point is hard to get and nodes have to make a long time commitment to accummulate the points.

\subsection{\stair\ Framework}
\begin{figure}[ht]
	\centering
	\includegraphics[width=0.7\linewidth]{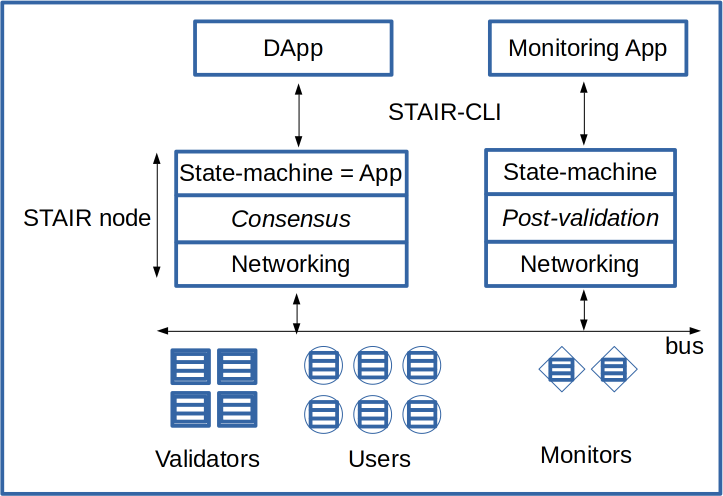}
	\caption{An overview of \stair\ framework}
	\label{fig:stairframework}
\end{figure}

Figure~\ref{fig:stairframework} depicts an overview of the \stair\ framework. A validating node (User or Validator) consists of three components: state machine, consensus and networking.  A DApp can communicate via a CLI. \stair\ framework supports auditing by permitting participants to join  in post-validation mode. An observer (or Monitor) node consists of state machine, post validation component and networking component.

\subsubsection{Stake-aware Gossip}

In \stair\ protocol, it aims for an asynchronous permissionless leaderless system. Each node can create and exchange messages to / from other nodes.

Like our StakeDag~\cite{lachesis01,stakedag}, each node batches client transactions into a new event block and stores in its own x-DAG. Each node exchanges its own blocks as well as the ones it received from other nodes asynchronously. The protocol uses the concept of distributed common knowledge together with network broadcasting to reach a consistent global view with high probability from its local view.

Unlike StakeDag~\cite{stakedag} and previous PoS protocols, \stair\ protocol encourages User nodes (low stake) to connect with Validators (high stake) in order to gain more validating power for User's new own blocks. Validators are also encouraged to connect to User nodes so as to validate the User's new blocks.
The stake-aware gossip will improve the fairness, safety and liveness of the protocol.

\subsubsection{Main procedure}
Algorithm~\ref{al:main} shows the main function serving as the entry point to launch \stair\ protocol. The main function consists of two main loops, which are they run asynchronously in parallel. The first loop attempts to request for new nodes from other nodes, to create new event blocks and then communicate about them with all other nodes. 
The second loop will accept any incoming sync request from other nodes. The node will retrieve updates from other nodes and will then send responses that consist of its known events.

\begin{algorithm}[H]
	\caption{\stair\ Main Function}\label{al:main}	
	\begin{algorithmic}[1]
		\State $\mathcal{U}$ $\leftarrow$ set of users ($t_i < U$).
		\State $\mathcal{V}$ $\leftarrow$ set of validators ($t_i \geq U$).
		\Function{Main Function}{$n_m$}
		\BState \emph{loop}:
		\State Let $\{n_i\}$ $\leftarrow$ $k$-PeerSelectionAlgo($n_m$, $\mathcal{U}$, $\mathcal{V}$)
		\State Sync request to each node in  $\{n_i\}$
		\State (SyncPeer) all known events to each node in $\{n_i\}$
		\State Create new event block: newBlock($n_m$, $\{n_i\}$)
		\State (SyncAll) Broadcast out the message		
		\State Update x-DAG $G$
		\State Call ComputeConsensus($G$)
		\BState \emph{loop}:
		\State Accepts sync request from a node
		\State Sync all known events by \stair\ protocol
		\EndFunction
	\end{algorithmic} 
\end{algorithm}

The $k$PeerSelectionAlgo takes as input the current set of users $U$ and the current set of validators $V$. There are two cases depending on the running node is a User or a Validator.
\begin{itemize}
	\item
For a User node, it will first select a Validator $v_i$ from the set $\mathcal{V}$ with a probability based on the stake of $v_i$. It will then select $k$-1 nodes randomly with a probability based on the their stakes.
\item For a Validator, it follows similar procedure like above. It will first select a User $v_i$ from the set $\mathcal{U}$ with a probability based on the stake of $v_i$. It will then select $k$-1 nodes randomly with a probability based on the their stakes.
\end{itemize}

\subsubsection{Stake-based Validation}
In \stair, a node must validate its current block and the received ones before it attempts to create or add a new event block into its local DAG. A node must validate its (own) new event block before it communicates the new block to other nodes.

A root is an important block that is used to compute the final consensus of the event blocks of the DAG.
Like StakeDag protocol, \stair\ procol takes the stakes of participants into account to compute the consensus of blocks. A stake number of a block is the sum of the validating power of the nodes whose roots can be reached from the block. When a block receives more than 2/3 of the entire validating power of the network, it becomes a root.

\subsubsection{Validation Score of a Block}

When an event block becomes a root, it receives the validating power of the root's creator. The validation score of a block is the total of all validating powers that the block has gained.
The validation score of a block $v_i$ $\in$ $G$ is denoted by $s(v_i)$.
If the \emph{validation score} of a block is greater than 2/3 of the total validating power, the block becomes a root. The weight of a root $r_i$ is denoted by $w(r_i)$, which is the weight $w_j$ of the creator node $j$ of the $r_i$.

\begin{figure}[t]
	\centering
	\includegraphics[width=0.4\linewidth]{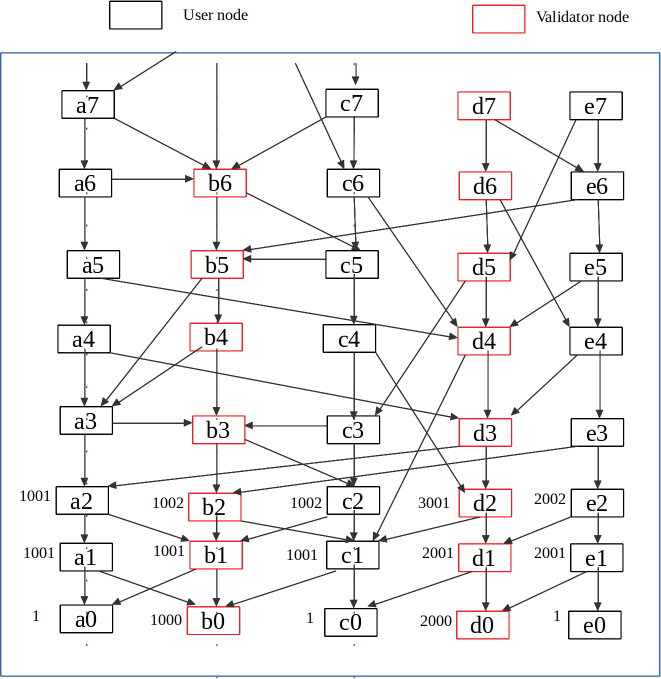}
	\caption{An example of an x-DAG chain in \stair. The validation scores of some selected blocks are shown.}
	\label{fig:stakedag-validatorscore}
\end{figure}
Figure \ref{fig:stakedag-validatorscore}
depicts an example of x-DAG. In this example, two validators (in Red) have validating power of 1000 and 2000 respectively, while users (in Black) have validating power of 1. First few event blocks are marked with their validating power. Leaf event blocks are special as each of them has a validation score equal to their creator's validating power.

\subsection{Checkpoint}
Like StakeDag, our \stair\ protocol improves validation procedure over the previous PoS approaches. \stair\ uses the following procedure based on the Casper model~\cite{buterin2018}. 
Our approach can be summarised as follows:
\begin{itemize}
	\item Every 100th frame is a checkpoint. Each block is 1-20 seconds and each frame is 1-10 minutes.
	\item Stakeholders can choose to make more deposits at each check point, if they want to become validators and earn more rewards. Validators can choose to exit, but cannot withdraw their deposits until three months later.
	\item A checkpoint is selected based on a consistent global history that is finalized with more than 2/3 of the validating power for the checkpoint. When a checkpoint is finalized, the transactions will not be reverted. 
	\item With asynchronous system model, validators are incentivized to coordinate with each other on which checkpoints the history should be updated.  Nodes gossip their latest local views to synchronize frames and accounts. Attackers may attempt double voting to target double spending attacks. Honest validators are incentivized to report such behaviors and burn the deposits of the attackers.
\end{itemize}

\subsection{Post validation}
Nodes do not require any tokens to join the network as observers. Observers can only perform post validation and cannot participate in onchain validation (voting). Intuitively, the observers have zero token and thus they have a validating power of zero. Thus, as far as trust and validating power is concerned, observers do not and cannot contribute to the online validation.

Observers can retrieve network events from users and validators. Like users and validators, observers can store a state machine in their local history, which is used to perform validation and check for integrity of the common ledger. 

The x-DAG is a persistent record, which stores who-validated-what. Post-validation is extremely useful, especially for auditing and for recovering from a crash. During post validation, observers can notify the network (delegates) about malicious records such as invalid blocks, illegal transactions, and undetected forks.

To encourage more post-validation, observers who are the first to detect malicious blocks and transactions, can be rewarded. The reward can be a portion of the burnt deposit of attackers. Alternatively, the observers can be rewarded with validation points (Saga points).
Observers are incentivised to post-check the integrity of the system. 

Observers can choose to upgrade to become Users and Validators by staking some amount of stake. If the observer gained some Saga points, these points will be remained after the upgrading.

\subsection{Consensus}

Nodes in \stair\ protocol have a local x-DAG. Updates on each node are synchronized to the all the nodes.
The consensus layer is similar to StakeDag~\cite{stakedag}. Each node computes Roots, Frames and Clothos to reach consensus on finalized event blocks, called Atroposes.

\newpage
\section{\stair\ Staking model}\label{se:staking}

This section presents our staking model for PoS DAG-based \stair\ protocols. We first give general definitions and variables of our staking model, and then present the model in details.

\subsection{Definitions and Variables}

Below are the general definitions and variables that are important in \stair\ protocol.

\textbf{General definitions}

\begin{longtable}{p{1cm} p{13cm}}
	$\mathcal{F}$ & denotes the network itself \\[2pt] 
	$SPV$          &"Special Purpose Vehicle" -- managing the collection of transaction fees and the payment of all rewards \\[2pt] 
	$FTM$          & main network token \\[2pt] 
\end{longtable}

\textbf{Accounts}

\begin{longtable}{p{1cm} p{1.2cm} p{13cm}}
	$\mathbb{U}$  & & set of all participant accounts in the network \\[2pt] 
	$\mathbb{A}$  & $\subset \mathbb{U}$ & accounts with a positive FTM token balance \\[2pt] 
	$\mathbb{S}$  & $\subseteq \mathbb{A}$ & accounts that have staked for validation (some of which may not actually be validating nodes) \\[2pt] 
	$\mathbb{V}$  & $\subseteq \mathbb{S}$ & validating accounts, corresponding to the set of the network's validating nodes
\end{longtable}

A participant with an account having a positive FTM token balance, say $i \in \mathbb{A}$, can join the network.
But an account $i$ in $\mathbb{S}$ may not participate in the protocol yet. Those who join the network belong to the set $\mathbb{V}$. 

\dfnn{Validating account}{A validating account is an account with a positive FTM token balance}.

Note that, the set of validating accounts $\mathbb{V}$ contain both users and validators, as presented in our general model in Section~\ref{se:model}.

\textbf{Network parameters subject to on-chain governance decisions}

\begin{longtable}{p{1cm} p{1.2cm} p{13cm}}
	$F$  & $3.175e9$ & total supply of FTM tokens \\[2pt] 
	$\lambda$ & $90$ & period in days after which validator staking must be renewed, to ensure activity \\[2pt] 
	$\varepsilon$ & 1 & minimum number of tokens that can be staked by an account for any purpose \\[2pt] 
	$\phi$    & 30\% & SPV commission on transaction fees \\[2pt] 
	$\mu$     & 15\% & validator commission on delegated tokens
\end{longtable}

\textbf{Tokens held and staked}

Unless otherwise specified, any mention of \textit{tokens} refers to FTM tokens.

\dfnn{Token helding}{The token helding  $t_i$ of an account is number of FTM tokens held by account $i \in \mathbb{A}$.}

\begin{longtable}{p{1cm} p{1.2cm} p{13cm}}
	$t_i$        &                 & number of FTM tokens held by account $i \in \mathbb{A}$ \\[2pt] 
	$t_i^{[x]}$      & $> \varepsilon$ & transaction-staked tokens by account $i$\\[2pt] 
	$t_i^{[d]}(s)$   & $> \varepsilon$ & tokens delegated by account $i$ to account $s \in \mathbb{S}$ \\[2pt] 
	$t^{[d]}(s)$     &                 & total of tokens delegated to account $s \in \mathbb{S}$ \\[2pt] 
	$t_i^{[d]}$      &                 & total of tokens delegated by account $i$ to accounts in $\mathbb{S}$ \\[2pt] 
	$t_i^{[s]}$      &                 & validation-staked tokens by account $i$ \\[2pt] 
\end{longtable}

The sum of tokens staked or delegated by an account $i \in \mathbb{A}$ cannot exceed the amount of tokens held:
$
t_i^{[x]} + t_i^{[s]} + \sum_{s \in \mathbb{S}} t_i^{[d]}(s)  \leq t_i
$.
The total amount of tokens delegated to an account $s \in \mathbb{S}$ is: $
t^{[d]}(s) = \sum_{i \in \mathbb{A}} t_i^{[d]}(s)
$. 
The total amount of tokens delegated by an account $i \in \mathbb{A}$ is:$
t_i^{[d]} = \sum_{s \in \mathbb{S}} t_i^{[d]}(s)
$

\subsection{Validating power}

We use a simple model of validating power, which is defined as the number of tokens held by an account.
The weight of an account $i \in \mathbb{A}$ is equal to its token holding $t_i$.

\dfnn{Validating power}{
	The validating power of a validator $v \in \mathbb{V}$ is defined as validator's weight. The weight of a validator is defined based on the token helding $t_i$.}

The validator's weight of node $n_i$ is denoted by $w_i$ or as a function $w(t_i)$. Like StakeDag, we can use simple model: $w_i$ = $t_i$. 
In \stair, we also propose a threshold value of trust ($U$) to separate between users and validators.
The validating power or weight $w_i$ is defined by: 

\[
w_i= 
\begin{cases}
U \times \lfloor \frac{t_i}{U} \rfloor,& \text{if } t_i \geq U\\
1,              & \text{otherwise}
\end{cases}
\]

\subsection{Block consensus and Rewards}
We then present notations and definitions in our model of rewards.

\dfnn{Validation score} {Validation score of a block is the total validating power that a given block can achieve from the validators $v \in \mathbb{V}$.}

\dfnn{Validating threshold}{Validating threshold is defined by $2/3$ of the validating power that is needed to confirm an event block to reach consensus.}

Below are the variables that define the block rewards and their contributions for participants in Fantom network.

\begin{longtable}{p{1.2cm} p{2.5cm} p{11cm}}
	$Z$      & 996,341,176 & total available block rewards of $FTM$, for distribution by the $SPV$ during the first 1460 days after mainnet launch\\[2pt] 
	$F_s$    &   & $FTM$ tokens held by the $SPV$ \\[2pt] 
	$F_c$    & $F - F_s$ & total circulating supply \\[2pt] 	
\end{longtable}

Block rewards will be distributed over 1460 days (4 years less one day) after launch, corresponding to $Z / 1460$ per day during that period: $R_b(d)$ = $682,425.46$ (during 1460 days after mainnet launch), or 0 (otherwise).


\subsection{Token Staking and Delegation}\label{se:stakingoverview}

FTM tokens will have multiple uses in the Fantom system. Participants can chose to stake or delegate tokens into their accounts. When staking or delegating, the validating power of a node is based on the number of FTM tokens held.
Three possible ways for staking to achieve wealth are given as follows.

\dfnn{Transaction staking}{Participants can gain more stakes or tokens via the transaction staking.}
Transaction submitters will gain transaction fees if the transactions are successfully validated and reach finality.
This style of staking helps increases transaction volume on the network. The more liquidity, the more transaction staking they can gain.

\dfnn{Validation staking}{By validating blocks, a participant can gain validation rewards. Honest participants can gain block rewards for  successfully validated blocks.}

With validation staking, one can join the network using a moderate hardware and a certain small amount of tokens. Once participating the network, the participant can achieve block rewards for blocks that they co-validated and gain transaction fees from transaction submitters for the successfully finalized transactions. The more stake they gain, the more validating power they will have and thus the more rewards they can receive as it is proportional to their validating power.

\dfnn{Validation delegation}{Validation deligation allows a participant to deligate all or part of their tokens to another participant(s). Deligating participants can gain a share of block rewards and transaction fees, based on the amount of delegated stake.}

Our model of delegation of stakes between participants allow various coordination amongst participants. Early stage participants can borrow stake from other participants to gain more rewards. Meanwhile, participants with large amount of stake can deligate their stakes to another participant while still earning some shared rewards.

Stakeholders will be able delegate a portion of their tokens to validating nodes. Validators will not be able to spend delegated tokens, which will remain secured in the stakeholder’s own address. Validators will receive a fixed proportion of the validator fees attributable to delegators.

\dfnn{Validation performance}
Participants are compared for their validation efforts. Uptimes and successful validation rates will result in more rewards. Delegators will be incentivised to choose nodes that have a high validator score, i.e. are honest and high performing.
Delegators can delegate their tokens for a maximum period of days, after which they need to re-delegate. The requirements to delegate are minimal:
\begin{itemize}
	\item \textit{Minimum number of tokens per deligation}: 1
	\item \textit{Minimum lock period}: 1 day
	\item \textit{Maximum number of delegations made by a user}: None
	\item \textit{Maximum number of delegated tokens  per validator}: 15 times of the validator's tokens.
\end{itemize}

\section{Discussions}\label{se:discuss}

This section presents discussions and benefits of our proposed DAG-based PoS approach.
We then compare the protocol fairness and security aspects with existing PoW and PoS blockchains. 

\subsection{Choices of $L$ and $U$}

The value of lower bound $L$=$1$ is used to separate between Observers and Users. The upper bound $U$=$1000$ separates between Users and Validators. Both $L$ and $U$ are configurable and will vary based on the community's decision.

\begin{figure}[ht]
	\centering
	\includegraphics[width=0.8\linewidth]{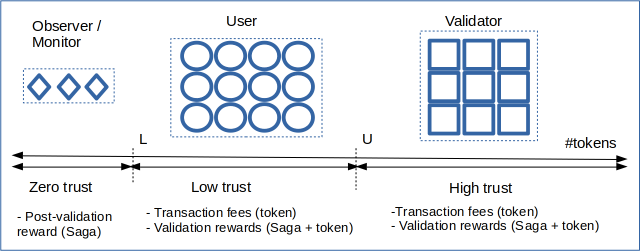}
	\caption{Token holding, trust levels and incentives of the three account types}
	\label{fig:stairaccounts}
\end{figure}

Figure~\ref{fig:stairaccounts} shows the differences between three different account types. The level of trust is proportional to the number of tokens. Observers have a trust level of zero and they have incentives to do post-validation for Saga points.
Users have low trust as their token holding varies from 1 to $U$, whereas Validators have high trust level with more than $U$ tokens in their account. Both Users and Validators have incentives to join the network to earn transaction fees and validation rewards.

Presumingly, there are a limited number of Observers and a small pool of Validators in the entire network. The number of Users may take a great part of the network.

\subsection{Alternative models for gossip and staking}
The principle that \stair\ protocol aims for is a more balanced chance for both Users and Validators. This is to improve liveness and safety of the network. In Section~\ref{se:model}, an example of stake-aware gossip and x-DAG construction is presented. An example of staking model is presented in Section~\ref{se:staking}.

Here, we present possible alternative model for gossip and staking. Specifically, the stake of a node is equal to the the token holding of the node, say $w_i = t_i$. The value of $U$ can be a predefined  constant, or can be selected based on the set of stakes $w_i$. Each User node randomly selects a Validator node in $\mathcal{V}$ and each Validator node randomly chooses a User node in $\mathcal{U}$. The probability of being selected is proportional to the value $w_i$ in the set, in both cases.
Alternatively, each Validator node can randomly select a node  from all nodes with a probability reversely proportional to the stake $w_i$.

\subsection{Post-validation and network load}
Observers may also consume network throughput from the validating nodes. 
The number of observers is expected to be relatively small.

However, when the number of observers is large, observers may not be allowed to synchronize and retrieve events directly from Users and Validators to avoid slowing down the network. Instead, we suggest to use a small pool of Moderator nodes that will retrieve updated events from Users and Validators. Observers will then synchronize with the Moderator nodes to download new updates and to perform post-validation.

\subsection{Protocol Fairness and Security}

Regarding the fairness, PoW protocol is fair because a miner with $p_i$ fraction of the total computational power can win the reward and create a block with the probability $p_i$.
PoS protocol is fair given that an individual node, who has $w_i$ fraction of the total stake or coins, can a new block with $w_i$ probability.
However, in PoS systems, initial holders of coins tend to keep their coins in their balance in order to gain more rewards.

Our \stair\ protocol is fair since every node has an equal chance to create an event block. Nodes in \stair\ can enter the network without an expensive hardware like in PoW. Further, any node in \stair\ protocol can create a new event block with the same propability, unlike stake-based probability of block creation in PoS blockchains.

Like a PoS blockchain system, it is a possible concern that the initial holders of coins will not have an incentive to release their coins to third parties, as the coin balance directly contributes to their wealth. Unlike PoS, that concern in \stair\ protocol is about the economic rewards a \stair\ node may get is proportional to the stake they possess after they successfully contribute to the validation of event blocks. 

\begin{table}[h]
	\centering
	\begin{tabular}{|l|c|c|c|c|}
		\hline
		Criterion & PoW & PoS & StakeDag & \stair \\\hline
		Block creation probability & $p_i /P$ & $w_i /W $ & $1/n$ & $1/n$ \\\hline
		Validation probability & $1/n$ & $w_i/W$ & $w_i/W$ & $1/n$ \\\hline
		Validation reward & $p_i /P$ & $w_i /W$ & $w_i /W$ & $\alpha_i w_i /W$ \\\hline
		Txn Reward & $p_i /P$ & $w_i /W$ & $1/n$ & $1/n$ \\\hline
		Performance reward & - & - & - & +$1$ Saga point \\
		\hline
	\end{tabular}
	\vspace{5pt}
	\caption{Comparison of PoW, PoS, StakeDag and \stair\ Protols}\label{tab:comp}
\end{table}

Table~\ref{tab:comp} gives a comparison of PoW, PoS, StakeDag and new \stair\ protocols.
Let $p_i$ denote the computation power of a node and $P$ denote the total computation power of the network. Let $w_i$ be the stake of a node and $W$ denote the total stake of the whole network. Let $\alpha_i$  denote the number of Saga points a node has been rewarded from successful validations of finalized blocks.

Remarkably, our \stair\ protocol is more intuitive because our reward model used in stake-based validation can lead to a more reliable and sustainable network.

For security, like StakeDag, \stair\ has less vulnerabilities than PoW, PoS and DPoS. \stair\ protocol is more secure than PoW and PoS.
For more details about vulnerabilities of PoW and PoS, see StakeDag~\cite{stakedag}.

\section{Conclusion}\label{se:con}
In this paper, we introduce a new stake-based consensus protocol, so-called \stair, for permissionless asynchronous trustless systems.
The new protocol is based on our StakeDag framework~\cite{stakedag}.
Specifically, we propose a new type of DAG, namely x-DAG or cross-DAG, to reach faster consensus. We have also presented a new staking model, which is simple. We also present our \stair\ framework with three different types of role: Users, Validators and Observers. Last but not least, we propose a new notion of Saga points to measure the validation performance between participants.
\stair\ leverages cross validation and post-checks to ensure integrity and sustainability of the whole network.

\clearpage
\section{Reference}\label{se:ref}

\renewcommand\refname{\vskip -1cm}
\bibliographystyle{abbrv}
\bibliography{LCA}

\end{document}